\documentclass[aps,preprint]{revtex4}%
\usepackage{amsfonts}
\usepackage{amsmath}
\usepackage{amssymb}
\usepackage{graphicx}%
\providecommand{\U}[1]{\protect\rule{.1in}{.1in}}

\begin{document}
\title[ ]{Conformal Symmetry and Planck's Thermal Radiation in Classical Physics}
\author{Timothy H. Boyer}
\affiliation{Department of Physics, City College of the City University of New York, New
York, New York 10031, U.S.A.}
\affiliation{email: tboyer@ccny.cuny.edu}
\keywords{}
\pacs{}

\begin{abstract}
Thermal radiation is discussed for a classical scalar field. \ Dilation and
acceleration appear in conformal symmetry, and so are treated in connection
with zero-point radiation and equilibrium thermal radiation. \ Classical
zero-point radiation is completely invariant under dilation in an inertial
frame. \ However, in a Rindler frame, zero-point radiation takes a different
form which is related to Planck's thermal spectrum with temperature $T=\hbar
a/\left[  2\pi ck_{B}\right]  $ where $a$ is the acceleration of the point.
\ Traditionally, zero-point radiation is said to take a thermal (Unruh) form,
in a Rindler frame. \ However, the classical viewpoint indicates that
zero-point radiation in a Rindler frame requires a divergent frequency
spectrum and so is not thermal radiation. \ The Unruh form suggests an
additional classical thermal spectrum satisfying the Stefan-Boltzmann
relation. \ 

\end{abstract}
\maketitle

\section{Introduction}

\subsection{The Puzzle of the Unruh Effect in Classical Physics}

The connection between the quantum Unruh effect and classical physics has been
a puzzle for some time. \ In 1980, an analysis was given of an apparently
classical aspect\cite{B42} of the Unruh effect.\cite{BD} \ When a point had an
acceleration $a$ through \textit{classical} zero-point radiation, it seemed to
experience a Planck-like correlation function with an apparent temperature $T$
given by $T=\hbar a/\left(  2\pi ck_{B}\right)  $. \ Several subsequent
articles\cite{bsev} seemed to strengthen this classical perspective.
\ However, gradually, it came to be believed, among those physicists
interested in classical electromagnetic theory, to be a mistaken impression.
\ In this article, we will give a new derivation of the Planck thermal
radiation spectrum based on ideas connected with conformal symmetry. \ We will
again\cite{B-FoP} suggest that, within classical physics, on acceleration
through classical zero-point radiation, one does \textit{not} find a thermal
spectrum. \ Rather, classical thermal radiation is present in a box only if
the thermal radiation exists in the inertial frame of the box before the
acceleration began.

The important idea given here is that, within classical physics, the Unruh
idea gives a suggestion toward the Planck thermal spectrum but does not give
thermal radiation. \ We need to add on an additional radiation spectrum in
order to arrive at the Planck thermal spectrum. \ We can regard the radiation
spectrum as corresponding to a physical acceleration in spacetime, but also
recognize that conformal symmetry plays a role in an inertial frame. \ We will
introduce an acceleration-like spectrum for the correlation function of a
classical scalar field in an inertial frame, and will show that this
correlation function corresponds to that arising from the Planck thermal
radiation spectrum.

\subsection{Classical Zero-point Radiation versus Thermal Radiation}

We will show that, within classical theory, zero-point radiation and thermal
radiation are quite different, yet are connected. \ Within classical
electromagnetic theory, zero-point radiation is the radiation which causes the
Casimir effect between two conducting parallel plates.\cite{Casimir} \ It is
also the radiation which provides the ground state of atoms and molecules
despite the acceleration of the electrons.\cite{acc} \ Classical zero-point
radiation is dilation \textit{in}variant and Lorentz \textit{in}variant.

On the other hand, thermal radiation has a preferred inertial frame, the
inertial frame of the containing box, and has exactly one variable parameter,
the temperature $T.$ \ Under a dilation, the temperature changes. \ Thus under
dilation, thermal radiation is \textit{co}variant, not \textit{in}variant. We
will explain this contrast when discussing various classical radiation
spectra. \ 

\section{The System: Classical Scalar Field}

Here we describe thermal radiation for a relativistic \textit{scalar} wave in
terms of classical relativistic physics including classical zero-point
radiation. \ In classical physics, the radiation and the associated spacetime
remain the same, independent of the coordinate system used to describe the
system. \ The coordinate system can always be changed, so long as the metric
is changed with it. \ 

A massless scalar wave in four spacetime dimensions has a Hamiltonian in an
inertial frame%
\begin{equation}
H=\frac{1}{2}\int d^{3}r\left[  \frac{1}{c^{2}}\left(  \frac{\partial\phi
}{\partial t}\right)  ^{2}+\left(  \frac{\partial\phi}{\partial x}\right)
^{2}+\left(  \frac{\partial\phi}{\partial y}\right)  ^{2}+\left(
\frac{\partial\phi}{\partial z}\right)  ^{2}\right]  . \label{H}%
\end{equation}
The equation of motion is
\begin{equation}
\frac{1}{c^{2}}\frac{\partial^{2}\phi}{\partial t^{2}}-\frac{\partial^{2}\phi
}{\partial\times^{2}}-\frac{\partial^{2}\phi}{\partial y^{2}}-\frac
{\partial^{2}\phi}{\partial z^{2}}=0.
\end{equation}

Except for the velocity constant $c$, the equation of motion has no preferred
length, no preferred time, and zero mass. \ Therefore we expect the function
$\phi\left(  ct,\mathbf{r}\right)  $ to satisfy conformal symmetry. A harmonic
massless scalar wave propagating in the direction $\widehat{k}$ of (angular)
frequency $\omega$ travels with the speed of light $c$ as
\begin{equation}
\phi_{\mathbf{k}}\left(  t,\mathbf{r}\right)  =A_{\mathbf{k}}\cos\left[
\mathbf{k\cdot r}-kct+\theta\left(  \mathbf{k}\right)  \right]  ,
\end{equation}
where $\left\vert \mathbf{k}\right\vert =\omega/c$, $A_{\mathbf{k}}$ is an
amplitude, and $\theta\left(  \mathbf{k}\right)  $ is a phase.

We will assume that the wave satisfies homogeneous Dirichlet boundary
conditions at the walls of a cubic box $V=L^{3}$. \ For a box of finite
dimension $L$, there is clearly a largest wavelength corresponding to a
standing wave. \ For a cubic box of side $L$, the full wave takes the form of
a linear combination of the modes,
\begin{equation}
\phi\left(  t,\mathbf{r}\right)  =\sum_{\mathbf{k}}A_{\mathbf{k}}\cos\left[
\mathbf{k\cdot r}-kct+\theta\left(  \mathbf{k}\right)  \right]  .
\end{equation}

We will assume that the wave field corresponds to random, isotropic radiation,
with the phase $\theta\left(  \mathbf{k}\right)  $ random on the interval
$(0,2\pi]$ with each vector $\mathbf{\ k}$ corresponding to a different
independent phase. \ However, we will take the box very large, and will often
invoke the infinite-$L$ limit, in which case we would write the field
$\phi\left(  t,\mathbf{r}\right)  $ as existing over all spacetime,
\begin{equation}
\phi\left(  t,\mathbf{r}\right)  =\int d^{3}kA\left(  k\right)  \cos\left[
\mathbf{k\cdot r}-kct+\theta\left(  \mathbf{k}\right)  \right]  .
\end{equation}

\section{Dilation Invariance of Classical Zero-Point Radiation}

\subsection{Invariance with an Undetermined Constant}

If we calculate the energy in a normal mode of frequency $\omega=ck$ in a
cubic box of side $L,$ we find from the Hamiltonian given in Eq. (\ref{H})
that the energy $U(\omega)$ of the wave changes with the amplitude of the
wave. \ If we want a dilation-invariant wave, \ we expect it to take the form%
\begin{equation}
U\left(  \omega\right)  =const\times\omega
\end{equation}
where the constant \textquotedblleft$const$\textquotedblright\ is independent
of the frequency $\omega.$ \ This follows since, in the conformal group, the
energy $U$ follows the same pattern as the frequency $\omega.$ Thus we have%
\begin{equation}
\mathbf{r}^{\prime}=\sigma\mathbf{r,}\text{ }t^{\prime}=\sigma t,~\omega
^{\prime}=\omega/\sigma,~U^{\prime}=U/\sigma
\end{equation}
where $\sigma$ is any arbitrary, positive number. \ We expect that the vacuum
in rectangular coordinates in Minkowski spacetime has no preferred point, or
length, or time, or energy. \ The energy expression for a normal mode will
also have no preferred energy.

\subsection{Casimir Force Gives the Scale of Energy}

The Casimir force between two uncharged, parallel, conducting plates suggests
an electromagnetic energy of the form
\begin{equation}
U^{zp}\left(  \omega\right)  =\left(  1/2\right)  \hbar\omega\label{Uzpw}%
\end{equation}
where $\hbar$ is a fundamental constant. \ We will adopt this same energy for
our scalar waves. \ It turns out that the value of $\hbar$ is the same number
as was first introduced by Planck\cite{P1899} in 1899 before the advent of
quantum theory. \ Some physicists are surprised, even offended, to find the
same constant $\hbar$ appearing in a purely classical theory.

In order for dilation-invariant zero-point radiation in a scalar classical
theory to give the energy in (\ref{Uzpw}), we must write the radiation field
as
\begin{equation}
\phi^{zp}\left(  ct,xy,z\right)  =\int d^{3}k\sqrt{\frac{\hbar c}{2\pi^{2}k}%
}\cos\left[  \mathbf{k\cdot r-}kct\mathbf{+}\theta\left(  \mathbf{k}\right)
\right]  , \label{phizpctx}%
\end{equation}
where the amplitude is $A\left(  k\right)  =\sqrt{\hbar c/\left(  2\pi
^{2}k\right)  }.$

\section{Two-Point Field Correlation Function for $\phi^{zp}$}

We will consider the two-point correlation function \ when discussing thermal
radiation, and so wish to calculated this same correlation function for the
zero-point field. \ For the dilation-invariant zero-point field $\phi^{zp}$,
the correlation function takes the form%

\begin{align}
&  \left\langle \phi^{zp}\left(  ct,x,y,z\right)  \phi^{zp}\left(
c\overline{t},\overline{x},\overline{y},\overline{z}\right)  \right\rangle
_{\theta}\nonumber\\
&  =\int d^{3}k\,\sqrt{\frac{\hbar c}{2\pi^{2}k}}\int d^{3}k^{\prime}%
\,\sqrt{\frac{\hbar c}{2\pi^{2}k^{\prime}}}\left\langle \cos\left[
\mathbf{k\cdot r}-kct+\theta\left(  \mathbf{k}\right)  \right]  \cos\left[
\mathbf{k}^{\prime}\mathbf{\cdot r}^{\prime}-k^{\prime}ct^{\prime}%
+\theta\left(  \mathbf{k}^{\prime}\right)  \right]  \right\rangle _{\theta
}\nonumber\\
&  =\int d^{3}k\sqrt{\frac{\hbar c}{2\pi^{2}k}}\int d^{3}k^{\prime}%
\,\sqrt{\frac{\hbar c}{2\pi^{2}k^{\prime}}}\cos\left[  \mathbf{k\cdot}\left(
\mathbf{r-}\overline{\mathbf{r}}\right)  -k\left(  ct-c\overline{t}\right)
\right]  \frac{1}{2}\delta^{3}\left(  \mathbf{k-k}^{\prime}\right) \nonumber\\
&  =\frac{1}{2}\int d^{3}k\,\frac{\hbar c}{2\pi^{2}k}\cos\left[
\mathbf{k\cdot}\left(  \mathbf{r-}\overline{\mathbf{r}}\right)  -k\left(
ct-c\overline{t}\right)  \right]  , \label{corrtx}%
\end{align}
where the random phases $\theta\left(  \mathbf{k}\right)  $ are summed over
as
\begin{equation}
\left\langle \cos\left[  \theta\left(  \mathbf{k}\right)  \right]  \cos\left[
\theta\left(  \mathbf{k}^{\prime}\right)  \right]  \right\rangle =\left\langle
\sin\left[  \theta\left(  \mathbf{k}\right)  \right]  \sin\left[
\theta\left(  \mathbf{k}^{\prime}\right)  \right]  \right\rangle =\left(
1/2\right)  \delta^{3}(\mathbf{k-k}^{\prime})
\end{equation}
and
\begin{equation}
\left\langle \cos\left[  \theta\left(  \mathbf{k}\right)  \right]  \sin\left[
\theta\left(  \mathbf{k}^{\prime}\right)  \right]  \right\rangle =0.
\end{equation}
Performing the angular integration, we find%

\begin{align}
&  \left\langle \phi^{zp}\left(  ct,x,y,z\right)  \phi^{zp}\left(
c\overline{t},\overline{x},\overline{y},\overline{z}\right)  \right\rangle
\nonumber\\
&  =\frac{\hbar c}{4\pi^{2}}\int_{0}^{\infty}dk\,k\left(  2\pi\right)
\int_{0}^{\pi}d\psi\sin\psi\cos\left[  k\left\vert \mathbf{r-}\overline
{\mathbf{r}}\right\vert \cos\psi-k\left(  ct-c\overline{t}\right)  \right]
\nonumber\\
&  =\frac{\hbar c}{2\pi\left\vert \mathbf{r-r}^{\prime}\right\vert }\int%
_{0}^{\infty}dk\left\{  \sin\left[  k\left\{  \left\vert \mathbf{r-}%
\overline{\mathbf{r}}\right\vert -\left(  ct-c\overline{t}\right)  \right\}
\right]  +\sin\left[  k\left\{  \left\vert \mathbf{r-}\overline{\mathbf{r}%
}\right\vert +\left(  ct-c\overline{t}\right)  \right\}  \right]  \right\}
\label{corrtx2}%
\end{align}
The expression we have found is finite at low frequencies, but is oscillating
ever more rapidly at high frequencies. \ In order to give the integral
convergence, we introduce a small cut off with constant $\epsilon.$ \ Then we
take $\epsilon\rightarrow0$ to find%

\begin{equation}
\int_{0}^{\infty}dk\sin\left[  kb\right]  \exp\left[  -k\epsilon\right]
=\frac{1}{2i}\left(  \frac{\exp\left[  k\left(  ib-\epsilon\right)  \right]
}{ib-\epsilon}-\frac{\exp\left[  k\left(  -ib-\epsilon\right)  \right]
}{-ib-\epsilon}\right)  _{0}^{\infty}\rightarrow\frac{1}{b}.
\end{equation}
Accordingly, the correlation function in Eq. . (\ref{corrtx2}) becomes%
\begin{align}
&  \left\langle \phi^{zp}\left(  ct,x,y,z\right)  \phi^{zp}\left(  ct^{\prime
},x^{\prime},y^{\prime},z^{\prime}\right)  \right\rangle \nonumber\\
&  =\frac{\hbar c}{2\pi\left\vert \mathbf{r-r}^{\prime}\right\vert }\left(
\frac{1}{\left\{  \left\vert \mathbf{r-}\overline{\mathbf{r}}\right\vert
-\left(  ct-c\overline{t}\right)  \right\}  }+\frac{1}{\left\{  \left\vert
\mathbf{r-}\overline{\mathbf{r}}\right\vert +\left(  ct-c\overline{t}\right)
\right\}  }\right)  \nonumber\\
&  =\frac{-\hbar c}{\pi}\left(  \frac{1}{\left(  ct-c\overline{t}\right)
^{2}-\left\vert \mathbf{r-}\overline{\mathbf{r}}\right\vert ^{2}}\right)
.\label{corrtx3}%
\end{align}

\subsection{Conformal \textit{In}variance of Classical Zero-Point Radiation}

We notice that this correlation function is dilation \textit{in}variant.
\ Under a dilation, we have $\phi^{zp\prime}=\phi^{zp}/\sigma,$ but then
\begin{align}
\left\langle \phi^{zp\prime}\left(  ct^{\prime},x^{\prime},y^{\prime
},z^{\prime}\right)  \phi^{zp\prime}\left(  c\overline{t}^{\prime}%
,\overline{x}^{\prime},\overline{y}^{\prime},\overline{z}^{\prime}\right)
\right\rangle  &  =\left\langle \phi^{zp}\left(  c\sigma t,\sigma x,\sigma
y,\sigma z\right)  \phi^{zp}\left(  c\sigma\overline{t},\sigma\overline
{x},\sigma\overline{y},\sigma\overline{z}\right)  \right\rangle \nonumber\\
&  =\frac{-\hbar c}{\pi}\left(  \frac{1}{\left(  c\sigma t-c\sigma\overline
{t}\right)  ^{2}-\left\vert \sigma\mathbf{r-}\sigma\overline{\mathbf{r}%
}\right\vert ^{2}}\right)  \nonumber\\
&  =\left(  \frac{1}{\sigma}\right)  ^{2}\left\langle \phi^{zp}\left(
ct,x,y,z\right)  \phi^{zp}\left(  c\overline{t},\overline{x},\overline
{y},\overline{z}\right)  \right\rangle .
\end{align}
Thus indeed we have $\phi^{zp\prime}=\phi^{zp}/\sigma$ making zero-point
radiation dilation invariant. \ 

\section{Thermal Radiation \ \ }

\subsection{Stefan-Boltzman Relation}

According to the Stefan-Boltzman relation in classical theory, thermal
radiation at temperature $T$ is a finite amount of energy $U^{T}$ in a
finite-volume container $V=L^{3}$ satisfying
\begin{equation}
U^{T}=a_{S}T^{4}L^{3},\label{SB}%
\end{equation}
where $a_{S}$ is a constant. \ This expression does \textit{not} give the
Casimir force, which may involve arbitrarily high frequencies. \ In order to
include both thermal and zero-point radiation, thermal radiation in classical
theory is pictured as involving a sum%
\begin{equation}
U^{total}\left(  \omega,T\right)  =U^{T}\left(  \omega,T\right)
+U^{zp}\left(  \omega\right)
\end{equation}
and
\begin{align}
&  \left\langle \phi^{total}\left(  ct,x,y,z\right)  \phi^{total}\left(
c\overline{t},\overline{x},\overline{y},\overline{z}\right)  \right\rangle
\nonumber\\
&  =\left\langle \phi^{T}\left(  ct,x,y,z\right)  \phi^{T}\left(
c\overline{t},\overline{x},\overline{y},\overline{z}\right)  \right\rangle
+\left\langle \phi^{zp}\left(  ct,x,y,z\right)  \phi^{zp}\left(  c\overline
{t},\overline{x},\overline{y},\overline{z}\right)  \right\rangle ,
\end{align}
where the finite density of thermal radiation is in addition to the zero-point
radiation and involves different random phases so that $\left\langle \phi
^{T}\left(  ct,x,y,z\right)  \phi^{zp}\left(  c\overline{t},\overline
{x},\overline{y},\overline{z}\right)  \right\rangle =0.$

\subsection{Puzzle of Thermal Radiation}

The spectrum of the finite amount of thermal radiation remained a mystery at
the end of the 19th century. \ At low frequencies, the Rayleigh-Jeans law held%
\begin{equation}
U^{RJ}\left(  ck,T\right)  =k_{B}T\text{ \ for \ }k_{B}T>>\left(  1/2\right)
\hbar\omega.
\end{equation}
But at very high frequencies, there was apparently no radiation at all, since
zero-point radiation was not considered.

In 1896, Wien showed that thermal radiation must follow the rule
\begin{equation}
U^{T}\left(  \omega,T\right)  =\omega F\left(  \omega/T\right)  ,
\end{equation}
where $F(\omega/T)$ was some unknown function. \ This rule was consistent with
dilation, since both $\omega$ and $T$ varied as $1/\sigma,$ but did not
indicate what function $F\left(  \omega/T\right)  $ was involved.

It was Planck's suggestion\cite{Planck} in 1900 that%
\begin{equation}
U^{T}\left(  ck,T\right)  =\frac{\hbar\omega}{\exp\left[  \hbar\omega/\left(
k_{B}T\right)  \right]  -1}\label{Plancks}%
\end{equation}
was the thermal radiation spectrum. \ Notice again, that the radiation at very
high frequencies goes to zero very rapidly; there is no zero-point radiation.
\ This expression has been accepted by many physicists as the full thermal
radiation energy. \ However, it does \textit{not} give the observed Casimir
forces. \ Also there has been some confusion as to the basis for this expression.

\subsection{Dilation in an Inertial Frame}

The Stefan-Boltzmann law (\ref{SB}) is consistent with the temperature
transforming as an energy under dilation, since%
\begin{equation}
U^{T\prime}\left(  ck^{\prime},T^{\prime}\right)  =U^{T}/\sigma=a_{S}\left(
\frac{T}{\sigma}\right)  ^{4}\left(  \sigma L\right)  ^{3}=\left(
a_{S}TL\right)  ^{4}/\sigma=\left(  1/\sigma\right)  U^{T}\left(  k,T\right)
.
\end{equation}
The Planck spectrum in Eq. . (\ref{Plancks}) is also consistent with this
scale invariance since%
\begin{align}
U^{T\prime}\left(  ck^{\prime},T^{\prime}\right)   &  =\left(  1/\sigma
\right)  U^{T}\left(  ck/\sigma,T/\sigma\right)  \nonumber\\
&  =\left(  1/\sigma\right)  \hbar\omega/\sigma\left[  \exp\left[  \left(
\hbar\omega/\sigma\right)  /\left(  k_{B}T/\sigma\right)  \right]  -1\right]
=\left(  1/\sigma\right)  U^{T}(ck^{\prime}T^{\prime}).
\end{align}

\section{Acceleration and the Unruh Effect}

\subsection{Acceleration Through Zero-Point Radiation}

In 1980 following the work on the Unruh effect in quantum field theory, an
analogous calculation\cite{B42} was done using classical physics. \ At that
time, it was suggested that on acceleration $a$ of a point through zero-point
radiation, one found the correlation function corresponding to the Planck
spectrum of thermal radiation with a temperature $T=\hbar a/\left(  2\pi
ck_{B}\right)  .$ \ The correlation function at a single point in space and
two different times follows from Eq. (\ref{corrtx3}) as
\begin{align}
&  \left\langle \phi^{zp}\left(  ct,x,y,z\right)  \phi^{zp}\left(
c\overline{t},x,y,z\right)  \right\rangle \nonumber\\
&  =\left\langle \phi^{zp}\left(  \xi\sinh\eta,\xi\cosh\eta,y,z\right)
\phi^{zp}\left(  \xi\sinh\overline{\eta},\xi\cosh\overline{\eta},y,z\right)
\right\rangle \nonumber\\
&  =\frac{\hbar c}{\pi}\frac{1}{\xi^{2}\left[  \left(  \cosh\eta
-\cosh\overline{\eta}\right)  ^{2}-\left(  \sinh\eta-\sinh\overline{\eta
}\right)  ^{2}\right]  }\nonumber\\
&  =\frac{\hbar c}{\pi}\frac{-1}{4\xi^{2}\sinh^{2}\left[  \left(
\eta-\overline{\eta}\right)  /2\right]  },\label{corRR}%
\end{align}
when the acceleration is taken in the $x$-direction and the associate
coordinates in the Rindler frame are $\eta$ and $\xi,$
\begin{equation}
ct=\xi\sinh\eta,~x=\xi\cosh\eta,\label{ctxeta}%
\end{equation}
while $y$ and $z$ are unchanged. \ The acceleration of the point is given by%
\begin{equation}
a=c^{2}/\xi.
\end{equation}

Notice that this correlation function (\ref{corRR}) still satisfies invariance
under dilation since $\eta$ is independent of the scaling parameter $\sigma$
while the coordinate $\xi$ scales just like time $t$ and space $x$.
\ Moreover, the \textit{spatial} correlation function at a single time $t$
still corresponds to classical zero-point radiation, not thermal radiation,
since the metric is now%
\begin{equation}
ds^{2}=\xi^{2}d\eta^{2}-d\xi^{2}-dy^{2}-dz^{2}.
\end{equation}
The correlation function in Eq. (\ref{corRR}) is exactly equal to that in Eq.
(\ref{corrtx3}). \ 

This situation corresponds to a Rindler frame where the time varies with
position, but the spatial distance does not vary. \ Locally, we can imagine
that we are living in a Rindler frame. \ Thus, while the clocks on the bottom
floor of a skyscraper run slower than the clocks at the top of the building,
all the spatial distances, such as the distance between floors, do \textit{not
}change. \ Whereas the separation between the floors is always the same as
measured in the comoving inertial reference frame, the time separation changes
between frames in second order in time. \ In a deep mine, clocks run slower
than on the surface of the earth. \ The slight time difference between the
floors is too small to be noticeable for a tall building. \ However, this
small difference becomes important for GPS calculations where satellites are
involved. \ 

We conclude that, within classical physics, thermal radiation is not
zero-point radiation as seen in a Rindler frame.

\section{Introduction of a Conformal-Invariant Constant $\zeta$ into the
Rindler Frame}

\subsection{Thermal Radiation and Conformal Symmetry\ }

Although thermal radiation is not zero-point radiation as seen in a Rindler
frame, the idea of acceleration being connected to a Planck-like spectrum
suggests that conformal symmetry is involved; accelerations appear in
conformal symmetry as do dilations. \ Zero-point radiation is smooth with no
preferred times or distances. \ A Rindler frame has no preferred lengths, but
it does have a preferred surface, $\xi=0,$ corresponding to the event horizon.

In previous work, it was suggested that while classical zero-point radiation
is dilation \textit{in}variant, thermal radiation is dilation \textit{co}%
variant with exactly one scaling parameter, the temperature $T$. \ Each person
living in a skyscraper can decide himself on the temperature in his own
apartment or on his own stove. \ People are not constrained regarding
temperature by their distance to the event horizon associated with the local
acceleration of their apartment in the Rindler frame. \ 

Suppose we take the Rindler expression in Eq. . (\ref{corRR}), but now
introduce a free parameter $\zeta$ so as to consider not $\xi$ but rather
$\xi/\zeta.$ \ In this case, we consider%
\begin{equation}
\left\langle \phi^{\zeta}\left(  ct,x,y,z\right)  \phi^{\zeta}\left(
c\overline{t},x,y,z\right)  \right\rangle =\frac{\hbar c}{4\pi}\left(
\frac{1}{\xi/\zeta}\right)  ^{2}\frac{-1}{\sinh^{2}\left[  \left(
c\overline{\tau}-c\tau\right)  /\left(  2\xi/\zeta\right)  \right]
}\label{phiGz}%
\end{equation}
in a Rindler frame. \ The event horizon is still located at $\xi=0$ as in Eq.
(\ref{ctxeta}), but the correlation function has changed from Eq.
(\ref{corRR}) over to Eq. (\ref{phiGz}). \ The presence of the parameter
$\zeta$ in the correlation function raises the possibility of equilibrium
thermal behavior which changes as $\xi$ changes in a Rindler frame. \ 

In a Rindler frame, as one goes to ever-larger values of the coordinate $\xi
$\ perpendicular to the event horizon, the temperature of the thermal
radiation decreases\cite{Tolman} as \
\begin{equation}
T\left(  \xi\right)  =\zeta/\xi.\label{Tolman}%
\end{equation}
where $\zeta$ is some \textit{temperature-related} parameter. \ 

The allowed normal modes for the box, the Dirichlet boundary conditions at the
walls of the box, and the distance to the event horizon all remain unchanged
by the introduction of the parameter $\zeta$. \ We still have a steady-state
distribution in the Rindler frame. \ For small separation in time,
$\tau-\overline{\tau},$ the constant $\zeta$ disappears, and the the
correlation function goes over to the zero-point radiation result in Eq.
(\ref{corrtx3}),%
\begin{align}
\left\langle \phi^{total}(ct,\xi,y,z)\phi^{total}\left(  ct^{\prime}%
,\xi,y,z\right)  \right\rangle  &  \rightarrow\frac{\hbar c}{4\pi}\left(
\frac{\zeta}{\xi}\right)  ^{2}\frac{-1}{\left[  \zeta\left(  c\overline{\tau
}-c\tau\right)  /\left(  2\xi\right)  \right]  ^{2}}\nonumber\\
&  =\frac{\hbar c}{\pi}\frac{-1}{\left(  c\overline{\tau}-c\tau\right)  ^{2}}.
\end{align}
Thus for large distances from the event horizon, the thermal radiation is only
at very low frequencies whereas the zero-point radiation remains even at high
frequencies. \ In the momentarily comoving inertial frame, the proper time
$\left(  c\overline{\tau}-c\tau\right)  $ agrees with the local time, $\left(
c\overline{\tau}-c\tau\right)  =c\overline{t}-ct.$

\subsection{Fourier Frequency Transform}

Although we have a correlation function for two different times given in Eq.
(\ref{phiGz}), we do not have the spectrum of radiation which would produce
such a correlation function in a Rindler frame. \ 

The Fourier transformation of the function $\omega\coth\left[  \alpha
\omega\right]  $ is\cite{G-R}%

\begin{align}
&  \int_{0}^{\infty}d\omega\left(  \omega\coth\left[  \frac{\xi}{\zeta}%
\frac{\pi\omega}{c}\right]  \right)  \cos\left[  \omega t\right] \nonumber\\
&  =\int_{0}^{\infty}d\omega\,\omega\cos\left[  \omega t\right]  +\int%
_{0}^{\infty}d\omega\left(  \frac{2\omega}{\exp\left[  2\pi\xi\omega/\left(
\zeta c\right)  \right]  -1}\right)  \cos\left[  \omega t\right] \nonumber\\
&  =-\frac{1}{t^{2}}+\left[  \frac{1}{t^{2}}-\left(  \frac{\zeta c}{2\xi
}\right)  ^{2}\text{csch}^{2}\left(  \frac{\zeta c}{2\xi}t\right)  \right]
=-\left(  \frac{\zeta c}{2\xi}\right)  ^{2}\text{csch}^{2}\left(  \frac{\zeta
c}{2\xi}t\right)  . \label{Fourierw}%
\end{align}
Here we have used the singular integral%
\begin{equation}
\int_{0}^{\infty}dk\,k\cos\left(  bk\right)  =\operatorname{Re}\lim
_{\lambda\rightarrow0}\int_{0}^{\infty}dk\,k\exp\left[  \left(  ib-\lambda
\right)  k\right]  =-1/b^{2}.
\end{equation}
We note that csch$\left(  x\right)  =1/\left[  \sinh\left(  x\right)  \right]
.$

Thus the correlation function for two different times at a single spatial
point can be rewritten from Eqs. (\ref{phiGz}) and (\ref{Fourierw}) as
\begin{align}
&  \left\langle \phi^{total}\left(  ct,\xi,y,z\right)  \phi^{total}\left(
ct^{\prime},\xi,y,z\right)  \right\rangle \nonumber\\
&  =\left(  \frac{-\hbar}{\pi c}\right)  \left\{  \left[  \left(  \frac{\zeta
c}{2\xi}\right)  ^{2}\text{csch}^{2}\left(  \frac{\zeta c}{2\xi}\left(
\tau-\overline{\tau}\right)  \right)  -\frac{1}{\left(  \tau-\overline{\tau
}\right)  ^{2}}\right]  +\frac{1}{\left(  \tau-\overline{\tau}\right)  ^{2}%
}\right\}  .
\end{align}
The part involving the square brackets is finite as $\left(  \tau
-\overline{\tau}\right)  \rightarrow0,$ and corresponds to the thermal
radiation, whereas the part $1/\left(  \tau-\overline{\tau}\right)  ^{2}$
diverges as $\left(  \tau-\overline{\tau}\right)  \rightarrow0$ and
corresponds to the zero-point radiation. \ The proper time $\tau$ at a single
spatial point $\xi,y,z$ is connected to the time $t$ at a single spatial point
$x,y,z$ in the mometarily comoving reference frame as in Eq. (\ref{ctxeta}).

\subsection{Connection of $\zeta/\xi$ with Temperature}

Now we want to leave out the divergent zero-point radiation and to integrate
over the finite \textit{thermal} part of the spectrum at a fixed coordinate
$\xi$ to obtain the connection of $\zeta/\xi$ with temperature $T$ as given in
the Stefan-Boltzmann law, $u^{T}\left(  T\right)  =a_{S}T^{4}$, where
$u^{T}(T)$ is the thermal energy density. \ Subtracting off the zero-point
radiation spectrum so as to leave only the thermal part of the spectrum, we
find\cite{G-R}%

\begin{align}
&  \int_{0}^{\infty}d\omega\frac{\omega^{2}}{\pi^{2}c^{3}}\left(  \frac{1}%
{2}\hbar\omega\coth\left[  \frac{\pi\xi}{\zeta c}\omega\right]  -\frac{1}%
{2}\hbar\omega\right) \nonumber\\
&  =\left(  \frac{\hbar c}{2\pi^{2}c^{4}}\right)  \left(  \frac{\zeta c}%
{\pi\xi}\right)  ^{4}\int_{0}^{\infty}du\,u^{3}\left(  \coth\left[  u\right]
-1\right) \nonumber\\
&  =\left(  \frac{\hbar c}{2\pi^{2}c^{4}}\right)  \left(  \frac{\zeta c}%
{\pi\xi}\right)  ^{4}\frac{1}{8}\frac{\pi^{4}}{15}=a_{S}T^{4}=\frac{\pi
^{2}k_{B}^{4}}{15\hbar^{3}c^{3}}T^{4},
\end{align}
where we have inserted the value for Stefan's constant $a_{S}=\left(  \pi
^{2}k_{B}^{4}\right)  /\left(  15\hbar^{3}c^{3}\right)  .$ \ Therefore the
connection with temperature is%

\begin{equation}
T=\frac{\left(  \zeta c\right)  \hbar}{2\pi\xi k_{B}}.\label{T}%
\end{equation}
We notice that the temperature $T$ in the Rindler frame varies with the
distance $\xi$ from the event horizon, exactly as given in Tolman's relation
(\ref{Tolman}). \ 

\subsection{The Thermal Radiation Spectrum}

The strictly thermal part of the radiation spectrum in a Rindler frame is
given by the result in Eq. (\ref{T}). \ However, there remains the zero-point
part which takes the Unruh form in a Rindler frame. \ Since both the thermal
radiation and the zero-point radiation are stationary in the Rindler frame, we
can bring the spectrum back to an inertial frame by stopping the acceleration
and joining the momentarily covariant inertial frame. \ But then we can
replace the expression $\zeta/\xi$ by the familiar temperature $T$. \ 

In an inertial frame, the zero-point radiation will take the familiar from
given in Eq. (\ref{corrtx3}). \ In an inertial frame, we would simply replace
$\zeta/\xi$ by $2\pi k_{B}T/\left(  \hbar c\right)  $. \ In an inertial frame
the correlation function for a scalar field would be
\begin{align}
&  \left\langle \phi^{total}\left(  ct,x,y,z\right)  \phi^{total}\left(
c\overline{t},x,y,z\right)  \right\rangle \nonumber\\
&  =\left(  \frac{\hbar c}{4\pi}\right)  \left(  \frac{2\pi k_{B}T}{\hbar
c}\right)  ^{2}\frac{-1}{\sinh^{2}\left[  \pi k_{B}T\left(  ct-c\overline
{t}\right)  /\left(  \hbar c\right)  \right]  }\nonumber\\
&  =\left(  \frac{\hbar c}{\pi c}\right)  \int_{0}^{\infty}d\omega\left(
\omega\coth\left[  \frac{\hbar\omega}{2k_{B}T}\right]  \right)  \cos\left[
\omega\left(  t-\overline{t}\right)  \right]  .
\end{align}
Comparing with Eq. (\ref{Plancks}), we have found exactly the Planck spectrum,
now including zero-point radiation,%

\begin{align}
U^{total}\left(  \omega,T\right)   &  =U^{T}\left(  \omega,T\right)
+U^{zp}\left(  \omega\right)  \nonumber\\
&  =\frac{\hbar\omega}{\exp\left[  \hbar\omega/\left(  k_{B}T\right)  \right]
-1}+\frac{1}{2}\hbar\omega=\frac{1}{2}\hbar\omega\coth\left[  \frac
{\hbar\omega}{2k_{B}T}\right]  .
\end{align}

\section{Comments on Thermal Radiation}

Many physicists assume that an acceleration $a$ though the vacuum produces a
thermal spectrum with temperature $T=\hbar a/\left(  2\pi ck_{B}\right)  $;
however, a classical analysis suggests otherwise. \  \ Within the classical
view, zero-point radiation takes a different form in a Rindler frame, and
thermal radiation must be above this zero-point radiation. \ Making an
alteration in the Unruh correlation function involving acceleration through
zero-point radiation in an Rindler frame, we are led to the Planck spectrum,
now including zero-point radiation, in an inertial or in a Rindler frame. \ We
emphasize that the presence of thermal radiation at a temperature above zero,
is required for a positive value of the constant $\zeta$. \ In a Rindler frame
at zero temperature, $\zeta/\xi=0$ for all coordinates $\xi,$ which means
$\zeta=0$. \ In this article, we have introduced an appropriate free parameter
$\zeta$ to account for thermal radiation. \ Within classical theory, the limit
$\zeta\rightarrow0$ corresponds to zero-point radiation which is always present.\

\end{document}